\documentstyle[11pt]{article}
\begin{document}

\newtheorem{defn}{Definition}
\newtheorem{intro}{Introduction}

\newcommand{\bs}{$\backslash$}
\newcommand{\comname}[1]{{\sf \bs #1}}
\newcommand{\envname}[1]{{\sf #1}}
\newcommand{\be}{\begin{equation}}
\newcommand{\ee}{\end{equation}}
\newcommand{\ben}{\begin{eqnarray}}
\newcommand{\een}{\end{eqnarray}}
\newcommand{\oz}{{\cal{A}}}
\newcommand{\moz}{{\cal{M}}}
\newcommand{\goz}{{\cal{G}}}
\newcommand{\hoz}{{\cal{H}}}
\newcommand{\doz}{{\cal{D}}}
\newcommand{\la}{\lambda}
\newcommand{\th}{\theta}
\newcommand{\no}{\noindent}
\newcommand{\hep}{C^{\infty}(V)}

\begin{titlepage}
\title{Electroweak Theory \\and\\ Noncommutative Geometry}
\author{ A. \c{C}atal-\"{O}zer\footnote{E.Mail:
acatal@newton.physics.metu.edu.tr} \, and
T. Dereli\footnote{E.Mail: tdereli@metu.edu.tr} \\
\\ {\small Department of Physics, Middle East Technical University}\\
{\small  06531 Ankara, Turkey}}
\date{ }
\maketitle

\begin{abstract}

\noindent The noncommutative generalisation of the standard
electroweak model due to Balakrishna, G\"{u}rsey and Wali
is formulated in terms of the derivations $Der_{2}(M_3)$ of
a three dimensional representation of the su(2) Lie algebra
of weak isospin. A light Higgs boson of mass about 130 GeV,
together with four very heavy scalar bosons are predicted.
\end{abstract}

\vskip 3cm

\noindent Talk given  at the {\bf  Second G\"{u}rsey Memorial
Conference: M-Theory and Dualities}, 19-24 June 2000, Bo\u{g}azi\c{c}i
University, \.{I}stanbul, Turkey.

\end{titlepage}


\section{Introduction}

The standard electroweak theory is one of the most remarkable achievements
of physics. In spite of its many successes it still cannot be considered
complete because the fundamental scalar field whose existence is essential
to the theory hasn't been observed.
The elusive Higgs particle is a direct consequence of the mass generation
mechanism that is induced by a quartic
scalar self-interaction potential with arbitrary coupling parameters.
Neither the strength of these couplings, nor the mass of the Higgs
boson can be determined within the standard model.
Therefore it is still worthwhile to look for a
deeper understanding of the related phenomena.
A popular approach to this problem involves introducing
higher dimensional space-times
with product topology so that the compact
extra dimensions may be related to the internal
symmetries of constituent particles.
In this picture the microstructure of spacetime  itself at
the Planck scales becomes a subject of speculations.
Several years ago a program
based on noncommutative geometries was started by
A. Connes \cite{connes1}.
The Connes-Lott  noncommutative electroweak
model \cite{connes2} and its later elaborations
\cite{kastler} assume essentially a four dimensional, two-sheeted
space-time and the Higgs scalars that are governed by a discrete
${\bf Z}_2$ symmetry relate different space-time sheets.
A little later Connes himself \cite{connes3} introduced
a notion of reality that helps to eliminate most
but not all unpleasant features of the original model.
There is a recent book \cite{madore} that may be consulted for
further literature.

The electroweak model based on a noncommutative geometry we study here
has a somewhat different structure as
the space-time degrees of freedom are
extended  by
matrices, which are supposed to describe the internal symmetries of
elementary particles.
It was introduced about ten years ago by Balakrishna, G\"{u}rsey and Wali
\cite{feza}. In this scheme,
the Higgs scalars arise naturally with the correct weak charge assignments
along with the gauge potentials  as part of the
connection and give
rise in the Yang-Mills action functional to a quartic Higgs potential
that appears already shifted
to a spontaneously broken symmetric phase.
This built-in mass generation mechanism involves a single
dimensionless coupling strength
 and two largely separated mass scales.
The subsequent renormalization group flow
from beyond Planck scales down to the electroweak scale
provides realistic values  for the
Weinberg mixing angle, the masses of the weak intermediate
bosons as well as the mass of a light Higgs boson.
The Higgs boson
 mass prediction of approximately $130 GeV/c^2$
comes remarkably  close to the energy range that generated much recent
 excitement among the Higgs searchers at CERN \cite{cern}.
We give below a short survey of the noncommutative structure
of the model and provide an analysis of the bosonic mass spectrum.
More details are found in our previous paper \cite{ac-td}.

\section{Generalities on Matrix Geometries}

 The basic data needed in
noncommutative geometry is the spectral triple or
the K-cycle ($\oz, \hoz, \doz$) were $\oz$ is an involutive
algebra of operators in Hilbert space $\hoz$ and $\doz$ is a
self-adjoint, unbounded operator in $\hoz$. In the
commutative case the spectral triple corresponds to the
 usual arena of doing
quantum field theory, namely,  $\oz = C^{\infty}(V)$ the algebra of smooth
functions on the space-time manifold $V$, $\hoz = L^{2}(V, S)$
the Hilbert space of $L^{2}$-spinors, and $\doz$ the Dirac
operator of the Levi-Civita spin connection.
Then a differential algebra from this spectral data is constructed
so that generalized connections and curvatures can be defined.
We should be able to follow a similar strategy in the noncommutative
case. To do that we first of all notice that
the group $Aut(\oz)$ of automorphisms of $\oz$ is isomorphic to
the group $Diff(V)$ of diffeomorphisms of $V$ in the usual
Riemannian case. The algebra describing the manifold is invariant
under automorphisms. Since the algebra is noncommutative, the
group $Aut(\oz)$ has a natural subgroup $Inn(\oz)$.
An automorphism $f$ is inner if
and only if it acts as conjugation of the elements of the algebra
by some unitary element $u \in \oz$, that is $f(a) = u a u^{*},
\forall a \in \oz$. All the other automorphisms are called outer
automorphisms and we can write $Aut(\oz) = Inn(\oz) \otimes
Out(\oz)$. In the noncommutative formulation of the standard model
inner and outer automorphisms appear as the transformations on the
space-time and the internal space,  respectively. Hence
the coordinate transformations on space-time get unified  with gauge
transformations.
We will extend the space-time degrees of freedom
by matrices in which case
the underlying $C^{*}$-algebra would be $\oz =
C^{\infty}(V) \otimes M_{n}({\bf C})$ where $
M_{n}({\bf C})$ is the algebra of $n \times n$ matrices.
This simple tensor product space is sufficient  for
our purpose of studying the bosonic sector of the electroweak
theory. The formalism developed here should be considered as a
particular case of a general theory and in this sense, it is
more appropriate  to call it ``matrix geometry'', rather
than ``noncommutative geometry'' ( see e.g. \cite{madore}).

 To construct the action we need four
 ingredients: differential forms on $\oz$, a Lie group G of
 ``internal symmetries'', a scalar product on the space of
 differential forms $\Omega^{\star}(\oz)$ and an
 invariant scalar product on the Lie algebra $\goz$ of the
 group G. The scalar product we introduce on
 $\Omega^{\star}(\oz)$ will be the substitute of the
 Dixmier trace. In fact
a manifold $V$ can be completely characterized by the commutative
$C^{*}$-algebra of continuous, complex-valued functions on it,
$C^{\infty}(V)$. In this matrix geometry version of noncommutative
geometry, our noncommutative manifold can be completely
characterized by replacing $C^{\infty}(V)$ by the algebra
$M_{n}({\bf C})$ which is again a $C^{*}-$algebra.
The Lie algebra of complex vector fields coincides with the Lie
algebra  of derivations  $Der(\hep)$. Similarly one can construct
the differential algebra of $M_{n}$ from the vector space
$Der(M_{n})$, space of all derivations of $M_{n}$.

\section{Balakrishna-G\"{u}rsey-Wali Model}

A choice of derivations  determines the model and the choice we
make is dictated by which symmetries we want unbroken at the end.
We use the Lie subalgebra $Der_{2}(M_{3})$ generated by a three
dimensional representation of su(2)  rather than the Lie algebra
$Der(M_{3})$ of all derivations of $M_{3}$. We have this extra
freedom because $Der(M_{n})$ is not a module over $M_{n}$. All
derivations of $M_{n}$ are inner. This means that every element X
of Der($M_{n}$) is of the form $X = adf$ for some $f$ in $M_{n}$.
In electroweak theory electromagnetic $U_{em}(1)$ whose generator
is $\tau_{0} + \tau_{3}$ will remain unbroken. Here $\tau_{0}$ is
identified with $Y+\frac{2}{3}$, where $Y$ is the hypercharge
$\frac{\tau_{8}}{\sqrt{3}}$, and $\tau_{3}$ and $\tau_{8}$ are the
usual Gell-Mann matrices. Among the above generators of $M_{3}$
only the generators of the U-spin subalgebra commute with
$\tau_{0} + \tau_{3}$ so we define our derivations as
\be\label{eqn2} e_{a}(f) = m_{a} [\lambda_{a}, f], \ \ \ \ \ f \in
M_{3} \ee

\noindent where a runs through the indices (+, -, 3) and
\ben \label{eqn3} \lambda_{\pm} & = & \frac{U_{\pm}}{\sqrt{2}} \quad
\quad , \quad \quad  \lambda_{3}  =  U_{3}  \een
with
\ben m_{\pm} & = & m  \quad \quad , \quad \quad  m_{3}  =  \frac{m^{2}}{M}.
\een
Here m and M are two mass scales
that have to be introduced into the theory to keep the dimensions
correct.
The vector space $Der_{2}(M_{n}$) has a Lie algebra structure given by
\be\label{eqn4} [e_{a}, e_{b}] = \sum_{c} \frac{m_{a} m_{b}}{m_{c}} \
C_{ab}^{~~c} \  e_{c} \ee \noindent and hence forms a Lie subalgebra
of $Der(M_{3})$. The structure constants $C_{ab}^{~~c}$ are
\be\label{eqn5} C_{+-}^{~~~3} = -C_{-+}^{~~~3} = 1, \ \ \ \ \ C_{3+}^{~~-}
= -C_{+3}^{~~-} = 1, \ \ \ \ \ C_{-3}^{~~+} = -C_{3-}^{~~+} = 1, \ee
\noindent with  all others equal to zero.

To define the algebra of forms $\Omega^{\ast}_{2}(M_{n})$ over
$M_{n}$ we first set  $\Omega^{0}_{2}(M_{n})$ equal to
$M_{n}$. Here the subindex 2 refers to the fact that we are using
the derivation algebra Der$_{2}(M_{3})$. Then we define $df$ for $f$
$\in M_{n}$ by the equation
\begin{equation}\label{eqn6}
  df(e_{a}) = e_{a}(f).
\end{equation}

\no The indices are going to be lowered and raised by the metric
\be\label{eqn8} g_{ab} = - Tr(\lambda_{a} \lambda_{b}). \ee

\no  Then we define the set of one forms $\Omega_{2}^{~1}(M_{3})$
to be the set of all elements of the form $f dg$ or $dg f$ with $f$ and
$g$ in $\oz$ subject to the relations $d(fg) = df g + f dg$. Because
of the noncommutativity \ \ $\lambda^{a}d\lambda^{b} \neq
d\lambda^{b}\lambda^{a}$,  the set $d\lambda^{a}$ is not a
convenient system of generators for $\Omega^{1}_{2}(M_{n})$. There
is a better set of generators completely characterized by the
equations
\begin{eqnarray}\label{eqn9}
  \theta_{\pm} (e_{\mp}) & = & 1 \quad , \quad  \theta_{\pm} (e_{3}) =
  0 , \\ \nonumber
  \theta_{3} (e_{\mp}) & = & 0 \quad , \quad
\theta_{3} (e_{3}) = 1.
\end{eqnarray}
These provide  the matrix analogue of the dual basis of 1-forms and
satisfy the structure equations

\be\label{eqn11} d\theta^{a} = \sum_{b,c} C^{~~a}_{bc} \ \frac{m_{b}
m_{c}}{m_{a}} \ \theta^{b} \wedge \theta^{c}. \ee
Note that all  $\theta^{a}$'s  commute with the elements of $M_{3}$.
From the generators $\theta^{a}$, we construct the 1-form
\begin{equation}\label{eqn15}
  \theta = - m_{a} \lambda_{a}\theta^{a}
\end{equation}
which is going to play an important role in the
definition of covariant derivatives and in the study of gauge
fields. It satisfies the zero-curvature condition
\begin{equation}\label{eqn20}
  d\theta + \theta^{2} = 0.
\end{equation}

We can generalize the above formalism to the case where the
algebra $\oz$ is the tensor product over the complex numbers of a
matrix algebras $M_{n}$ and the commutative algebra
$C^{\infty}(V)$, $\oz = M_{n}({\bf C}) \otimes C^{\infty}(V).$
This will be the underlying algebra in our study of the bosonic
sector of the electroweak theory. Let us choose a basis
$\theta^{\alpha}_{\beta} \ dx^{\beta}$ of $\Omega^{1}(V)$ over V
and let $e_{a}$ be the Pfaffian derivations dual to
$\theta^{\alpha}$. Set $i = (\alpha, a), 1 \  \leq i  \ \leq \ 4 +
3 = 7 $ and introduce $\theta^{i} = (\theta^{\alpha}, \theta^{a})$
as generators of $\Omega^{1}(\oz)$ as a left or right $\oz$-module
and $e_{i} = (e_{\alpha}, e_{a})$ as a basis of Der$_{2}(\oz)$ as
a direct sum
\begin{equation}\label{eqn12}
  \Omega^{1}({\mathcal{A}}) = \Omega^{1}_{h} \oplus \Omega^{1}_{v}
\end{equation}
where
\begin{equation}\label{eqn13}
  \Omega^{1}_{h} = \Omega^{1}(V) \otimes M_{n} \ \ \ \ \ , \ \ \ \
  \  \Omega^{1}_{v} = {\mathcal{C^{\infty}}}(V) \otimes \Omega^{1}(M_{n}).
\end{equation}
The exterior derivative $df$ of an element $f$ of $\mathcal{A}$ is
can be written as the sum horizontal and
vertical parts as
\begin{equation}\label{eqn14}
  df = d_{h}f + d_{v}f.
\end{equation}
In what follows we will work in Cartesian coordinates, that is we
take $e_{\alpha} = \partial_{\alpha}$ and $\th^{\alpha} =
dx^{\alpha}$. Hence we have \be\label{eqn31} d_{h} = dx^{\alpha}
\partial_{\alpha}. \ee

The gauge potential, which is an element of $\Omega^{1}(V)$ for a
trivial U(1)-bundle can be generalised to the noncommutative case
as an anti-Hermitian element of $\Omega^{1}(\oz)$. Let $\omega$ be
such an element of $\Omega^{1}(\oz)$. We can write it then as
\be\label{eqn17} \omega = A + \theta + \Phi \ee \noindent where
\ben\label{eqn18} A & = & - ig A_{\alpha} \theta^{\alpha} \quad
 \in \Omega^{1}_{h}(\oz)\\ \nonumber \Phi & = & g \phi_{a}
\theta_{a} \quad   \in \Omega^{1}_{v}(\oz) \een
\noindent and $\theta$ is as given by  ($\ref{eqn15}$).
 Note that $\omega$ is an anti-Hermitian
element of $\Omega^{1}({\mathcal{A}})$, that is, $\omega +
\omega^{\ast} = 0$ as it should be.
$g$ is the coupling
constant of the theory. $\phi_{a}$ are interpreted as the Higgs
fields. The gauge transformations of the trivial U(1)-bundle over V are
the unitary elements of $C^{\infty}(V)$. In analogy, we will
choose the group of local gauge transformations as the group of
unitary elements ${\cal{U}}$ of $\oz$, that is the group of
invertible elements of $u \in \oz$ satisfying $u u^{*} = 1$. Here
* is the *-product induced in $\oz$ and $\oz$ is considered as the set of functions
on V with values in $GL_{n}$.
The curvature 2-form $\Omega$ and the field strength $F$ are
defined as usual:
\begin{equation}\label{eqn23}
  \Omega = d\omega + \omega^{2} \quad \quad , \quad \quad   F = d_{h}A + A^{2}.
\end{equation}
In terms of components and with
\begin{equation}\label{eqn24}
  \Omega = \frac{1}{2} \ \Omega_{ij} \theta^{i} \wedge \theta^{j}
\quad , \quad  F = \frac{1}{2} F_{\alpha \beta} \theta^{\alpha}
  \wedge \theta^{\beta}
\end{equation}
we find
\begin{eqnarray}\label{eqn25}
\Omega_{\alpha \beta} & = &  F_{\alpha \beta} ,
\\ \nonumber \Omega_{\alpha a} & = &  g {\cal{D}}_{\alpha} \phi_{a} =
 g ( \partial_{\alpha} \phi_{a} - ig \ [A_{\alpha},
\phi_{a}] ) ,
\\ \nonumber  & & \\ \nonumber \Omega_{ab} & = & g^{2} [\phi_{a},
\phi_{b}] - 
g \sum_{c} \frac{m_{a} m_{b}}{m_{c}} \ C_{ab}^{~~c} \ \phi_{c}.
\end{eqnarray}

We can now write down the usual gauge
invariant Yang-Mills Lagrangian density 4-form \be\label{eqn26}
{\cal{L}} = -\frac{1}{2 g^{2}} \ Tr(\Omega_{ij} \Omega^{ij}). \ee
\no In terms of the components of $\Omega$ it  becomes
\be\label{eqn27} {\cal{L}} = -\frac{1}{2 g^{2}} \ Tr(F_{\alpha
\beta} F^{\alpha \beta}) + Tr({\cal{D}}_{\alpha} \phi_{a} \
{\cal{D}}^{\alpha} \phi^{a}) - V(\phi) \ee \no where the Higgs
potential
\be\label{eqn28} V(\phi) = - \frac{1}{2} Tr(\Omega_{ab} \Omega^{ab}).
\ee

\no From the form of $\Omega_{ab}$ in $(\ref{eqn25})$ we see that
$V(\phi)$ vanishes for values \be\label{eqn29} \phi_{a} = 0, \ \ \
\ \ \ \ \phi_{a} = \frac{m_{a}}{g} \ \la_{a}. \ee The first choice
corresponds to a symmetric vacuum,  while the spontaneously broken
symmetric phase obtains for the second vacuum configuration above.
In the latter case, the second term on the right hand side of
$(\ref{eqn27})$ becomes \be\label{eqn30} \frac{1}{g^{2}} \
Tr([A_{\alpha}, m_{a} \la_{a}] [A^{\alpha}, m_{a} \la^{a}]) \ee
\no which  is quadratic in gauge potentials and hence it
gives mass to  vector bosons. This means we have a naturally
built-in Higgs mechanism.

\section{Bosonic Mass Spectrum}

To study the scalar masses it is convenient to write the
three independent Higgs fields as follows:

\be\label{eqn32} \phi_{+} = \frac{H^{\dagger}}{\sqrt{2}}, \ \ \ \
\ \phi_{-} = \frac{H}{\sqrt{2}}, \ \ \ \ \ \phi_{3} = \triangle +
\frac{m^{2}}{2Mg} \ (2\tau_{0} - 1) \ee \no By using the metric
components $(\ref{eqn8})$ we see \be\label{eqn33} \phi^{+} = -2
\phi_{-},  \ \ \ \ \phi^{-} = -2  \phi_{+}, \ \ \ \ \phi^{3} = -2
\phi_{3} \ee
where
\begin{eqnarray}\label{eqn34}
 H & = & H_{+} V_{+} + H_{0} U_{+}  , \\ \nonumber
\triangle & = & \frac{1}{2} \ (\triangle_{0} \lambda_{0} +
\triangle_{a} \lambda_{a}).
\end{eqnarray}
\no For the gauge potentials we  write \be\label{eqn35} A  =  -
ig A_{\mu} dx^{\mu} = - ig \frac{1}{2} \ (B_{\mu} \lambda_{0} +
W_{\mu a} \lambda_{a}) dx^{\mu} \ee
$B$ and $W$'s are going describe  the gauge bosons,
while $H$'s and $\triangle$'s the scalar Higgs bosons.
Using the field components above we can write the connection
1-form  explicitly:

\begin{eqnarray}\label{connection}
\omega & = & A + \frac{g}{\sqrt{2}} \ H \theta_{-} +
\frac{g}{\sqrt{2}} \  H^{\ast} \theta_{+} + g \triangle \theta_{3}
\\ & & - \frac{m}{\sqrt{2}} \ U_{+} \theta_{-} -
\frac{m}{\sqrt{2}} \ U_{-} \theta_{+} + \frac{m^{2}}{4M} \
(\lambda_{0} + \lambda_{3}) \theta_{3} . \nonumber
\end{eqnarray}
The next step is to construct the corresponding curvature 2-form $\Omega$
with  components

\begin{eqnarray}\label{eqn36}
\Omega_{\mu \nu} & = &  F_{\mu \nu}  , \\ \nonumber \Omega_{\mu +} &
= & \frac{g}{\sqrt{2}} \ {\cal{D}}_{\mu} H , \ \ \ \ \ \ \ \
\Omega_{\mu - } = \Omega^{\ast}_{\mu +} , \\ \nonumber \Omega_{\mu
3} & = & g {\cal{D}}_{\mu} \triangle
\end{eqnarray}
where
\begin{equation}\label{eqn37}
  {\cal{D}}_{\mu} = d_{h} -i g [A_{\mu}, ]
\end{equation}
and the remaining three terms are
\begin{eqnarray}\label{eqn38}
\Omega_{+ -} & = & \frac{g^{2}}{2} \ [H, H^{\ast}] - g M \triangle
- m^{2} \lambda_{0} + \frac{m^{2}}{2} , \\ \nonumber \Omega_{+ 3} &
= & - \frac{g^{2}}{\sqrt{2}} \ \triangle H \ \ \ \ \ \ \ \Omega_{3
-} = \Omega_{+ 3}^{\ast}
\end{eqnarray}
We write out the Lagrangian density and read
 the Higgs potential
\begin{eqnarray}\label{eqn39}
  \frac{1}{g^{2}} \ V(H, \triangle) & = & \frac{1}{8} \ \left[H^{\dagger}H
  - \frac{m^{2}}{g^{2}}\right]^{2} \\ \nonumber
  & & + \frac{1}{4} \ \left[\frac{1}{2} \ H^{\dagger}H - \frac{M}{g} \
  \triangle_{0} - \frac{m^{2}}{g^{2}}\right]^{2} \\ \nonumber
  & & + \frac{1}{4} \ \left[\frac{1}{2} \ H^{\dagger} \sigma_{a} H -
  \frac{M}{g} \ \triangle_{a}\right]^{2} \\ \nonumber
  & & + \frac{1}{8} \ H^{\dagger} (\triangle_{0} + \triangle_{a}
  \sigma_{a})^{2} H .
\end{eqnarray}
\no Here  H is written as a two-column vector with entries $H_{+}$
and $H_{0}$ and $\sigma_{a}$ are the Pauli spin matrices.
The vacuum configuration can be found either directly from the
minimum of the above potential (which is a sum of squares) or from
$(\ref{eqn29})$ to be

\begin{equation}\label{eqn40}
  H_{0} = \frac{m}{g}  \ \ \ \ H_{+} = 0  \ \ \ \ \triangle_{0} =
  \triangle_{3} = - \frac{m^{2}}{2Mg}  \ \ \ \ \triangle_{1,2} = 0
\end{equation}
where only the electromagnetism survives symmetry breaking.

The mass spectrum of the model is as follows: The $W$ and $Z$
bosons have masses $\frac{m}{\sqrt{2}} \ \sqrt{1 +
\frac{m^{2}}{2M^{2}}}$ and $m$ respectively as can be calculated
from ($\ref{eqn30}$). To determine the
mass spectrum of the Higgs sector we  first write down the
linearized field equations for the Higgs fields and then
diagonalize the mass matrix \cite{ac-td}. What we have is five heavy scalars
with masses converging to $\sqrt{2} M$ in the limit $M \gg m$,
three zero-mass scalars referring to Goldstone modes which would
be absorbed by weak intermediate bosons to become massive, and
one light Higgs boson with mass $\sqrt{2} m$.

These mass relations are valid at the mass scale M, but we must
 predict the mass values at the electroweak scale $E_{Z} \sim m $.
This is done by
considering the renormalization group flow of the coupling
constants $g, g^{'}$ and the Higgs self-coupling constant
$\lambda$ down from the scale $M$ to scale $m$. The relations
$\lambda = \frac{g^{2}}{4}$ and $g = g^{'}$ imposed at the mass
scale $M$ will let us predict the Higgs masses. The relevant renormalisation
group equations are \cite{schrempp}
\begin{equation}\label{eqn1ren}
  16 \pi^{2} \frac{dg}{dt} = - \frac{19}{6} g^{3}
\end{equation}
\begin{equation}\label{eqn2ren}
16 \pi^{2} \frac{dg'}{dt} =  \frac{41}{6} g'^{3}
\end{equation}
\begin{equation}\label{eqn3ren}
  16 \pi^{2} \frac{d\lambda}{dt} = 24 \lambda^{2} - 3 \lambda (3
  g^{2} + g'^{2}) + \frac{3}{8}  \ [2 g^{4} + (g^{2} + g'^{2})^{2}]
\end{equation}
We first solve ($\ref{eqn1ren}$) and ($\ref{eqn2ren}$)
at  arbitrary scales $\mu$
subject to the
above constraints at the mass scale $M$.
Then we fix the measured values of $g$
and $g'$ at the electroweak  scale $\mu = E_{Z}=91 GeV$ which are $g(E_{Z}) =
0.4234$ and $g'(E_{Z}) = 0.1278$.
This requires  $M \sim 5 \times 10^{20} GeV$.
The remaining equation
($\ref{eqn3ren}$) can now be solved numerically by feeding in the
solutions of equations  ($\ref{eqn1ren}$) and
($\ref{eqn2ren}$), leading to the result  $\lambda(E_{Z}) = 0.14$.
Finally we  use  the standard model relation
\begin{equation}\label{eqn6ren}
  \frac{m_{H}^{2}(\mu)}{m_{Z}^{2}(\mu)} = \frac{8
  \lambda(\mu)}{g^{2}(\mu) + g'^{2}(\mu)}
\end{equation}
and noting that at scales $M$,  $m_{H} = m_{Z} = \sqrt{2} m$
and $ g^{2}(M) = g'^{2}(M) = 4 \lambda (M) = 0.49$
determine the Higgs boson mass at the electroweak scales.
To the extent that radiative corrections can be neglected
 $\cite{sirlin}$,
 we find
\be m_{H}(E_{Z}) \sim 130GeV. \ee

\section{Concluding Comments}

The noncommutative electroweak model of Balakrishna, G\"{u}rsey and
Wali is minimal in its construction and the bosonic
mass spectrum it predicts is realistic. The model could be generalised
in several directions.
One line of development would concern coupling of spinorial matter and
to seek supersymmetrization \cite{ellis}.
On the other hand it is emphasised very often in the literature
\cite{chamseddine,connes4} that
approaches based on noncommutative geometries
not only yield  promising reformulations of the standard model but
would also be relevant to gravitation.
The invariant tensor formulation of the
model given here is especially well suited for both of these lines of
investigation.

\newpage

\section{Acknowledgement}

We are grateful to Professor K. C. Wali for encouragement and
discussions.

\end{document}